\documentclass{jpsj-suppl}
\usepackage{txfonts,url,colortbl,arydshln,hhline} 
\makeatletter
\def\@cline#1-#2\@nil{%
  \noalign{\vskip-\arrayrulewidth}
  \omit
  \@multicnt#1%
  \advance\@multispan\m@ne
  \ifnum\@multicnt=\@ne\@firstofone{&\omit}\fi
  \@multicnt#2%
  \advance\@multicnt-#1%
  \advance\@multispan\@ne
  {\CT@arc@\leaders\hrule\@height\arrayrulewidth\hfill}%
  \cr}
 \makeatletter
\newcommand{\VLINE}{\multicolumn{1}{>{\centering\arraybackslash}p{\mysize} !{\color{white}\vline}}}
\newcommand{\DCX}{\cellcolor[gray]{0.8}}
\newcommand{\DCXobs}{\cellcolor[gray]{0.55}}
\newcommand{\STABLEa}[1]{\VLINE{\strut \cellcolor{black}\color[gray]{1.0} #1}}
\newcommand{\STABLEb}[1]{\rule{0pt}{\myheight}\strut \cellcolor{black}\color[gray]{1.0} #1}
\title{Search for Tetraneutron by Pion Double Charge Exchange Reaction at J-PARC}

\author{Hiroyuki \textsc{Fujioka}$^{1}$, Tomokazu \textsc{Fukuda}$^{2}$, Toru \textsc{Harada}$^{2}$, Emiko \textsc{Hiyama}$^{3}$,
Kenta \textsc{Itahashi}$^{3}$, Shunsuke \textsc{Kanatsuki}$^{1}$, Tomofumi \textsc{Nagae}$^{1}$, Takuya \textsc{Nanamura}$^{1}$, Takahiro \textsc{Nishi}$^{3}$}

\inst{$^{1}$Department of Physics, Kyoto University, Kyoto 606-8502, Japan \\
$^{2}$Department of Engineering Science, Osaka Electro-Communication University, Neyagawa, Osaka 572-8530, Japan \\
$^{3}$RIKEN Nishina Center, Wako, Saitama 351-0198, Japan}
\email{fujioka@scphys.kyoto-u.ac.jp}

\recdate{September 1, 2016}

\abst{Tetraneutron ($^4\mathrm{n}$) has come back in the limelight, because of recent
observation of a candidate resonant state at RIBF.
We propose to investigate the pion double charge exchange (DCX) reaction, i.e. ${}^4\mathrm{He}(\pi^-,\pi^+)$, as an alternative way to populate tetraneutron. An intense $\pi^-$ beam with the kinetic energy of $\sim 850\,\mathrm{MeV}$, much higher than that in past experiments at LAMPF and TRIUMF, will open up a possibility to improve the experimental sensitivity of the formation cross section, 
which will be much smaller than hitherto known DCX cross sections such as $^9\mathrm{Be}(\pi^-,\pi^+){}^9\mathrm{He}\,\mbox{(g.s.)}$.}
\kword{tetraneutron, pion double charge exchange reaction, J-PARC}

\begin{document}
\maketitle

\section{Introduction}
It remains an open question for more than half a century
whether a four-neutron system, i.e.~tetraneutron, as either a bound state or a resonance, 
exists or not~\cite{Kezerashvili}.
While various experiments had searched for a signature of bound states by various approaches,
recent \textit{ab-initio} calculations~\cite{Pieper,Hiyama} based on current understanding
of nuclear forces have ruled out the existence of the bound tetraneutron state.
However, recent observation of candidates of a resonant state with a narrow width 
by Kisamori \textit{et al.} at RIBF~\cite{SHARAQ}
has drawn renewed attention in studies of not only few-body systems in nuclear physics~\cite{Hiyama}
but also the equation of state of neutron stars~\cite{Nature}.

From the experimental side, 
a further investigation, including an independent approach with a completely different reaction,
 is needed in order to establish the existence of a tetraneutron state.
Charge-neutral tetraneutron may be produced by a double charge exchange (DCX) reaction on $^4\mathrm{He}$.
Indeed, the experiment at RIBF adopted the $^4\mathrm{He}({}^8\mathrm{He},{}^8\mathrm{Be})$ reaction, which was never studied in the past. 
The obtained energy is $0.83\pm 0.65\,\mbox{(stat.)} \pm 1.25\,\mbox{(syst.)}\,\mathrm{MeV}$ relative to the $4n$ decay threshold, and the upper limit of the width is $2.6\,\mathrm{MeV}$ in FWHM.
Motivated by their finding, we propose to use a pion DCX reaction,  $^4\mathrm{He}(\pi^-,\pi^+)$. 
Both the DCX reactions have a small momentum transfer for a zero-degree measurement,
in contrast to multi-nucleon transfer reactions such as $^7\mathrm{Li}({}^7\mathrm{Li},{}^{10}\mathrm{C})$~\cite{Aleksandrov}.

In 1980's two experiments to search for a bound tetraneutron
in the $^4\mathrm{He}(\pi^-,\pi^+)$ reaction were conducted
at LAMPF~\cite{Ungar} and TRIUMF~\cite{Gorringe},
but neither could find a signature of the bound state.
The incident beam energies were 165 and $80\,\mathrm{MeV}$, respectively.
While the LAMPF experiment measured $\pi^+$ scattered at $0^\circ$,
a time projection chamber covering $50^\circ$ and $130^\circ$ was utilized 
for the detection of $\pi^+$ at TRIUMF.
It should be noted that the pion DCX reaction can populate
neutron-rich (or proton-rich) nuclei in general (Fig.~\ref{nuclearchart}),
such as $^{18}\mathrm{C}$ (from $^{18}\mathrm{O}$)~\cite{Seth78}
and $^{9}\mathrm{He}$ (from $^{9}\mathrm{Be}$)~\cite{Seth87}.
For a comprehensive review, please refer to Refs.~\cite{Clement,Johnson}.

\begingroup
\hdashlinewidth=1mm
\hdashlinegap=0.5mm
\newlength{\mysize}
\newlength{\myheight}
\newlength{\raiseheight}
\setlength{\mysize}{0.8cm}
\setlength{\myheight}{0.78cm}
\setlength{\arrayrulewidth}{0.5pt}
\tabcolsep = 1pt
\setlength{\raiseheight}{.5\myheight}
\addtolength{\raiseheight}{-0.5em}
\begin{figure}
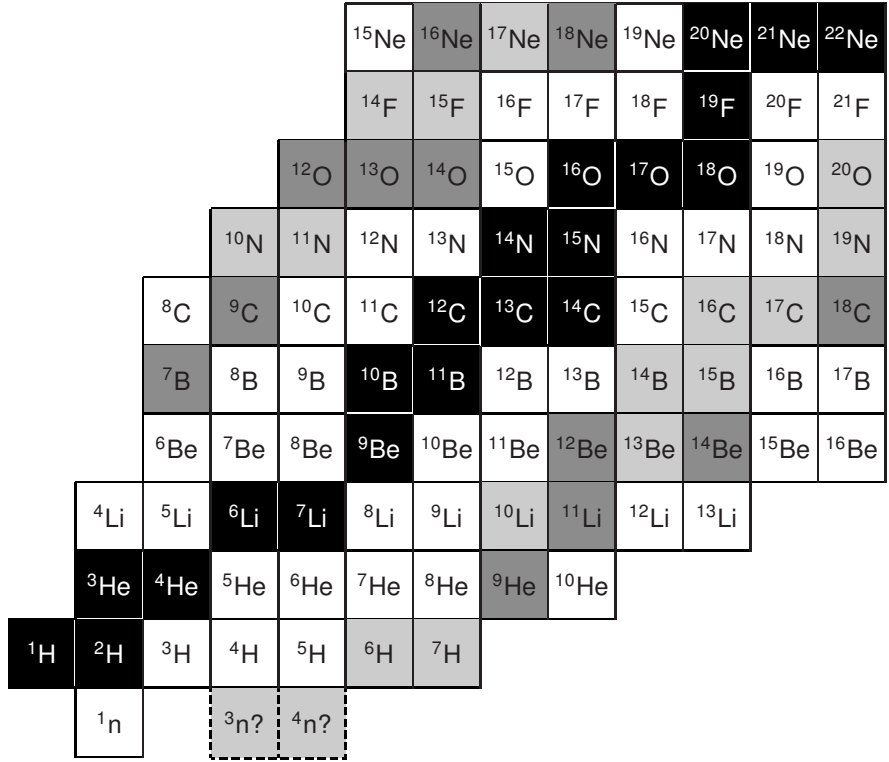

\begin{center}
\small
\sffamily
\begin{tabular}{*{13}{|>{\centering\arraybackslash}p{\mysize}}|}
\cline{6-13}
\multicolumn{5}{c|}{}&\raisebox{\raiseheight}{$\mathsf{^{15} Ne}$}&\raisebox{\raiseheight}{$\DCXobs\mathsf{^{16}Ne}$}&\raisebox{\raiseheight}{\DCX$\mathsf{^{17}Ne}$}&\raisebox{\raiseheight}{\DCXobs$\mathsf{^{18}Ne}$}&\raisebox{\raiseheight}{$\mathsf{^{19}Ne}$}&\STABLEa{\raisebox{\raiseheight}{{$\mathsf{^{20}Ne}$}}}&\STABLEa{\raisebox{\raiseheight}{$\mathsf{^{21}Ne}$}}&\STABLEb{\raisebox{\raiseheight}{$\mathsf{^{22}Ne}$}}\\
\cline{6-10}
\arrayrulecolor{white} \cline{11-11} \arrayrulecolor{black}
\cline{12-13}
\multicolumn{5}{c|}{}&\raisebox{\raiseheight}{\DCX$\mathsf{^{14}F}$}&\raisebox{\raiseheight}{\DCX$\mathsf{^{15}F}$}&\raisebox{\raiseheight}{$\mathsf{^{16}F}$}&\raisebox{\raiseheight}{$\mathsf{^{17}F}$}&\raisebox{\raiseheight}{$\mathsf{^{18}F}$}&\STABLEb{\raisebox{\raiseheight}{$\mathsf{^{19}F}$}}&\raisebox{\raiseheight}{$\mathsf{^{20}F}$}&\raisebox{\raiseheight}{$\mathsf{^{21}F}$}\\
\cline{5-10}
\arrayrulecolor{white} \cline{11-11} \arrayrulecolor{black}
\cline{12-13}
\multicolumn{4}{c|}{}&\raisebox{\raiseheight}{\DCXobs$\mathsf{^{12}O}$}&\raisebox{\raiseheight}{\DCXobs$\mathsf{^{13}O}$}&\raisebox{\raiseheight}{\DCXobs$\mathsf{^{14}O}$}&\raisebox{\raiseheight}{$\mathsf{^{15}O}$}&\STABLEa{\raisebox{\raiseheight}{$\mathsf{^{16}O}$}}&\STABLEa{\raisebox{\raiseheight}{$\mathsf{^{17}O}$}}&\STABLEb{\raisebox{\raiseheight}{$\mathsf{^{18}O}$}}&\raisebox{\raiseheight}{$\mathsf{^{19}O}$}&\raisebox{\raiseheight}{\DCX$\mathsf{^{20}O}$}\\
\cline{4-8}
\arrayrulecolor{white} \cline{9-9} \arrayrulecolor{black}
\cline{10-13}
\multicolumn{3}{c|}{}&\raisebox{\raiseheight}{\DCX$\mathsf{^{10}N}$}&\raisebox{\raiseheight}{\DCX$\mathsf{^{11}N}$}&\raisebox{\raiseheight}{$\mathsf{^{12}N}$}&\raisebox{\raiseheight}{$\mathsf{^{13}N}$}&\STABLEa{\raisebox{\raiseheight}{$\mathsf{^{14}N}$}}&\STABLEb{\raisebox{\raiseheight}{$\mathsf{^{15}N}$}}&\raisebox{\raiseheight}{$\mathsf{^{16}N}$}&\raisebox{\raiseheight}{$\mathsf{^{17}N}$}&\raisebox{\raiseheight}{$\mathsf{^{18}N}$}&\raisebox{\raiseheight}{\DCX$\mathsf{^{19}N}$}\\
\cline{3-7}
\arrayrulecolor{white} \cline{8-9} \arrayrulecolor{black}
\cline{10-13}
\multicolumn{2}{c|}{}&\raisebox{\raiseheight}{$\mathsf{^{8}C}$}&\raisebox{\raiseheight}{\DCXobs$\mathsf{^{9}C}$}&\raisebox{\raiseheight}{$\mathsf{^{10}C}$}&\raisebox{\raiseheight}{$\mathsf{^{11}C}$}&\STABLEa{\raisebox{\raiseheight}{$\mathsf{^{12}C}$}}&\STABLEa{\raisebox{\raiseheight}{$\mathsf{^{13}C}$}}&\STABLEb{\raisebox{\raiseheight}{$\mathsf{^{14}C}$}}&\raisebox{\raiseheight}{$\mathsf{^{15}C}$}&\raisebox{\raiseheight}{\DCX$\mathsf{^{16}C}$}&\raisebox{\raiseheight}{\DCX$\mathsf{^{17}C}$}&\raisebox{\raiseheight}{\DCXobs$\mathsf{^{18}C}$}\\
\cline{3-6}
\arrayrulecolor{white} \cline{7-7} \arrayrulecolor{black}
\cline{8-13}
\multicolumn{2}{c|}{}&\raisebox{\raiseheight}{\DCXobs$\mathsf{^{7}B}$}&\raisebox{\raiseheight}{$\mathsf{^{8}B}$}&\raisebox{\raiseheight}{$\mathsf{^{9}B}$}&\STABLEa{\raisebox{\raiseheight}{$\mathsf{^{10}B}$}}&\STABLEb{\raisebox{\raiseheight}{$\mathsf{^{11}B}$}}&\raisebox{\raiseheight}{$\mathsf{^{12}B}$}&\raisebox{\raiseheight}{$\mathsf{^{13}B}$}&\raisebox{\raiseheight}{\DCX$\mathsf{^{14}B}$}&\raisebox{\raiseheight}{\DCX$\mathsf{^{15}B}$}&\raisebox{\raiseheight}{$\mathsf{^{16}B}$}&\raisebox{\raiseheight}{$\mathsf{^{17}B}$}\\
\cline{3-13}
\arrayrulecolor{white} \cline{6-6} \arrayrulecolor{black}
\cline{7-13}
\multicolumn{2}{c|}{}&\raisebox{\raiseheight}{$\mathsf{^{6}Be}$}&\raisebox{\raiseheight}{$\mathsf{^{7}Be}$}&\raisebox{\raiseheight}{$\mathsf{^{8}Be}$}&\STABLEb{\raisebox{\raiseheight}{$\mathsf{^{9}Be}$}}&\raisebox{\raiseheight}{$\mathsf{^{10}Be}$}&\raisebox{\raiseheight}{$\mathsf{^{11}Be}$}&\raisebox{\raiseheight}{\DCXobs$\mathsf{^{12}Be}$}&\raisebox{\raiseheight}{\DCX$\mathsf{^{13}Be}$}&\raisebox{\raiseheight}{\DCXobs$\mathsf{^{14}Be}$}&\raisebox{\raiseheight}{$\mathsf{^{15}Be}$}&\raisebox{\raiseheight}{$\mathsf{^{16}Be}$}\\
\cline{2-13}
\multicolumn{1}{c|}{}&\raisebox{\raiseheight}{$\mathsf{^{4}Li}$}&\raisebox{\raiseheight}{$\mathsf{^{5}Li}$}&\STABLEa{\raisebox{\raiseheight}{$\mathsf{^{6}Li}$}}&\STABLEb{\raisebox{\raiseheight}{$\mathsf{^{7}Li}$}}&\raisebox{\raiseheight}{$\mathsf{^{8}Li}$}&\raisebox{\raiseheight}{$\mathsf{^{9}Li}$}&\raisebox{\raiseheight}{\DCX$\mathsf{^{10}Li}$}&\raisebox{\raiseheight}{\DCXobs$\mathsf{^{11}Li}$}&\raisebox{\raiseheight}{$\mathsf{^{12}Li}$}&\raisebox{\raiseheight}{$\mathsf{^{13}Li}$}\\
\cline{2-11}
\multicolumn{1}{c|}{}&\STABLEa{\raisebox{\raiseheight}{$\mathsf{^{3}He}$}}&\STABLEb{\raisebox{\raiseheight}{$\mathsf{^{4}He}$}}&\raisebox{\raiseheight}{$\mathsf{^{5}He}$}&\raisebox{\raiseheight}{$\mathsf{^{6}He}$}&\raisebox{\raiseheight}{$\mathsf{^{7}He}$}&\raisebox{\raiseheight}{$\mathsf{^{8}He}$}&\raisebox{\raiseheight}{\DCXobs$\mathsf{^{9}He}$}&\raisebox{\raiseheight}{$\mathsf{^{10}He}$}\\
\cline{1-1}
\arrayrulecolor{white} \cline{2-2} \arrayrulecolor{black}
\cline{3-9}
\STABLEa{\raisebox{\raiseheight}{$\mathsf{^{1}H}$}}&\STABLEb{\raisebox{\raiseheight}{$\mathsf{^{2}H}$}}&\raisebox{\raiseheight}{$\mathsf{^{3}H}$}&\raisebox{\raiseheight}{$\mathsf{^{4}H}$}&\raisebox{\raiseheight}{$\mathsf{^{5}H}$}&\raisebox{\raiseheight}{\DCX$\mathsf{^{6}H}$}&\raisebox{\raiseheight}{\DCX$\mathsf{^{7}H}$}\\
\cline{1-7}
\multicolumn{1}{c|}{}&\raisebox{\raiseheight}{$\mathsf{^{1}n}$}&\multicolumn{1}{c:}{}&\multicolumn{1}{c:}{\raisebox{\raiseheight}{\DCX$\mathsf{^{3}n}$?}}&\multicolumn{1}{c:}{\rule{0pt}{\mysize}\raisebox{\raiseheight}{\DCX$\mathsf{^{4}n}$?}}\\
\cline{2-2}\cdashline{4-5}
\end{tabular}
\end{center}
\caption{A part of the nuclear chart ($Z\le 10$ and $N\le 12$). Stable nuclei and long-lived $^{14}\mathrm{C}$, which was used as a target in past pion DCX measurements, are represented by black squares. Gray squares correspond to nuclides accessible by the $(\pi^\pm,\pi^\mp)$ reaction.  Nuclides observed in pion DCX reactions~\cite{TUNL} are highlighted in dark grey.}
\label{nuclearchart}
\end{figure}%
\endgroup

We aim to investigate pion DCX reactions
at much higher energy above the so-called $\Delta$ resonance region,
unexplored not only for tetraneutron but also for any other nuclides.
An intense secondary pion beam with the momentum $<2\,\mathrm{GeV/}c$ is available
at J-PARC Hadron Experimental Facility.
Due to a longer lifetime of pions in the laboratory frame, 
the purity of the incident $\pi^-$ beam and scattered $\pi^+$ can be improved,
which will contribute to a reduction of unphysical backgrounds in the region of interest.
In the past experiments, several events remained in the bound region, 
which had deteriorated the experimental sensitivity for a possible bound state.
An irrelevant background aside from the $4n$ continuum should be as small as possible,
no matter whether tetraneutron is a bound state or a resonant state.

Firstly, we will study an analog transition of $^{18}\mathrm{O}(\pi^+,\pi^-){}^{18}\mathrm{Ne}\,\mbox{(g.s.)}$, whose cross section would be much larger than that for a non-analog transition such as the $^{16}\mathrm{O}(\pi^+,\pi^-){}^{16}\mathrm{Ne}\,\mbox{(g.s.)}$ and $^4\mathrm{He}(\pi^-,\pi^+){}^4\mathrm{n}$ reactions, so as to examine the energy dependence of the formation cross section and establish the experimental procedure for a DCX measurement with a high-energy pion beam. In future, we will perform the spectroscopy of the $^4\mathrm{He}(\pi^-,\pi^+){}^4\mathrm{n}$ with a higher-intensity pion beam at the High-Intensity High-Resolution (HIHR) beamline to be constructed inside an extended Hadron Experimental Facility, which is in a planning stage~\cite{Noumi}.

\section{Analog-transition measurement at the K1.8 beamline}
The $^{18}\mathrm{O}(\pi^+,\pi^-){}^{18}\mathrm{Ne}\,\mathrm{(g.s.)}$ reaction
is one of the analog transitions,
since the final state is the double isobaric analog state (DIAS) with the isospin 1, the same as the ground state of the target.
Stable nuclides with the isospin 1 are scarce; the $^{18}\mathrm{O}$ nuclide is the lightest among them. 
The analog transition in $^{18}\mathrm{O}$ (as well as long-lived $^{14}\mathrm{O}$)
 at the incident energy up to $\sim 500\,\mathrm{MeV}$ was already investigated \cite{Williams},
 and the energy dependence was compared with that for a non-analog transition, $^{16}\mathrm{O}(\pi^+,\pi^-){}^{16}\mathrm{Ne}\,\mathrm{(g.s.)}$~\cite{Beatty}. 
They differ from each other by a factor of $\approx 20$ at the incident energy between 350 and $500\,\mathrm{MeV}$.

As for $>500\,\mathrm{MeV}$, only a theoretical calculation for the $^{18}\mathrm{O}(\pi^+,\pi^-){}^{18}\mathrm{Ne}\,\mathrm{(g.s.)}$ reaction exists~\cite{Oset}.
According to it, the forward differential cross section will decrease above $500\,\mathrm{MeV}$,
and reach a local minimum at $700\,\mathrm{MeV}$ and again increase up to $850\,\mathrm{MeV}$,
which corresponds to the third resonance region.

We plan to measure the forward differential cross section of the $^{18}\mathrm{O}(\pi^+,\pi^-){}^{18}\mathrm{Ne}\,\mathrm{(g.s.)}$ reaction~\cite{LoI_S2S} with the S-2S spectrometer~\cite{Kanatsuki} to be installed at the K1.8 beamline. 
The main purpose of the new spectrometer is a high-resolution $\Xi$-hypernuclear spectroscopy via the $^{12}\mathrm{C}(K^-,K^+){}^{12}_{\ \Xi}\mathrm{Be}$ reaction~\cite{Kanatsuki, Nagae}. By scanning the beam energy (momentum) around $850\,\mathrm{MeV}$ ($980\,\mathrm{MeV}/c$), the energy dependence can be compared with the theoretical calculation.
Moreover, owing to a relatively high yield ($\approx 400$ counts per day with a $10^7$/spill $\pi^+$ beam impinging on a $2\,\mathrm{g/cm^2}$-thick $\mathrm{{}^{18}O}$-enriched water target), a variety of studies, e.g. optics study and background reduction, will be feasible.

In order to realize the measurement, we need some modification of the experimental setup,
such as detectors for particle identification, currently optimized for the $(K^-,K^+)$ reaction.
A detailed study is in progress.

\section{Tetraneutron search at the HIHR beamline}
The LAMPF experiment deduced the upper limit of the differential cross section of a bound tetraneutron to be $22\,\mathrm{nb/sr}$ at the $1\sigma$ confidence level~\cite{Ungar}.
Taking into account theoretically-calculated energy dependence for the analog transition~\cite{Oset},
it is reasonable to aim at a sensitivity 
of $1\,\mathrm{nb/sr}$ in our measurement at $850\,\mathrm{MeV}$.
It is three orders of magnitude smaller than the expected cross section ($\sim 1\,\mathrm{\mu b/sr}$)
for the $^{18}\mathrm{O}(\pi^+,\pi^-){}^{18}\mathrm{Ne}\,\mathrm{(g.s.)}$ reaction 
discussed in the previous section.

By the way, a plan to extend the current Hadron Experimental Facility, 
and to construct new beamlines together with additional secondary-particle production targets,
is under discussion~\cite{Noumi}. 
The High-Intensity High-Resolution (HIHR) beamline, where the flagship experiment will be a high-precision $\Lambda$-hypernuclear spectroscopy with the $(\pi^\pm,K^+)$ reaction,
will be suited for tetraneutron search~\cite{LoI_HIHR}, too. 
According to the Sanford-Wang formula, the beam intensity will be $1.6\times 10^8$ per spill. At the same time, high resolution is expected thanks to dispersion matching between the beamline and the spectrometer.
If a $2.0\,\mathrm{g/cm^2}$-thick liquid $^{4}\mathrm{He}$ target is employed,
the yield of a tetraneutron state amounts to $97\,\mathrm{\mbox{events}/(2\,\mbox{weeks})/(nb/sr)}$. It is worth stressing that a high-resolution measurement will play an important role in determining the decay width in case of a resonant tetraneutron state, as well as improving the signal-to-noise ratio.

Last but not least, prior to the construction of the HIHR beamline,
a pilot measurement at the existing K1.8 beamline may be possible.
Even if only the upper limit of the formation cross section is determined experimentally,
the strength of the $4n$ continuum in the $^4\mathrm{He}(\pi^-,\pi^+)$ reaction will give a hint about 
more realistic experimental plan at the HIHR beamline.

\section{Conclusion and Outlook}
Stimulated by observation of resonant tetraneutron states at RIBF,
a pion DCX measurement at the incident energy around $850\,\mathrm{MeV}$ 
will be investigated at J-PARC for the first time.
Compared with the past experiments at LAMPF and TRIUMF,
high purity, high statistics, and high resolution will be achieved.

Lastly, we would like to point out a relevance to neutron-rich $\Lambda$ hypernuclei,
which are produced in the $(\pi^-,K^+)$ reaction.
While a recent experiment at J-PARC did not observe a peak structure for the ${}^6_\Lambda\mathrm{H}$ hypernucleus~\cite{Sugimura}, 
studies of the two kinds of DCX reactions using pion beams with similar energies may shed further light on underlying reaction mechanisms~\cite{Harada}.


\begin{thebibliography}{99}
\bibitem{Kezerashvili}R.Ya.~Kezerashvili: arXiv:1608.00169 [nucl-th], and references therein.
\bibitem{Pieper} S.C.~Pieper: Phys. Rev. Lett. \textbf{90} (2003) 252501.
\bibitem{Hiyama} E. Hiyama, R. Lazauskas, J. Carbonell, and M. Kamimura: Phys. Rev. C \textbf{93} (2016) 044004.
\bibitem{SHARAQ} K. Kisamori et al.: Phys. Rev. Lett. \textbf{116} (2016) 052501.
\bibitem{Nature}C.A. Bertulani and V.~Zelevinsky: Nature \textbf{532} (2016) 448.
\bibitem{Aleksandrov} D.V. Aleksandrov \textit{et al.}:  JETP Lett. \textbf{81} (2005) 43.
\bibitem{Ungar} J.E. Ungar et al.: Phy. Lett. \textbf{B144} (1984) 333.
\bibitem{Gorringe} T.P. Gorringe et al.: Phys. Rev. C \textbf{40} (1989) 2390.
\bibitem{Seth78} K.K.~Seth et al.: Phys Rev. Lett. \textbf{41} (1978) 1589.
\bibitem{Seth87} K.K.~Seth et al.: Phys Rev. Lett. \textbf{58} (1987) 1930.
\bibitem{Clement} H. Clement: Prog. Part. Nucl. Phys. \textbf{29} (1992) 175.
\bibitem{Johnson} M.B. Johnson and C.L. Morris: Annu. Rev. Nucl. Part. Sci. \textbf{43} (1993) 165.
\bibitem{TUNL} TUNL Nuclear Data Evaluation (\url{http://www.tunl.duke.edu/nucldata/})
\bibitem{Noumi}H.~Noumi, JPS Conf. Proc. \textbf{13} (2017) 010017.
\bibitem{Williams} A.L. Williams et al.: Phys. Lett. \textbf{B216} (1989) 11.
\bibitem{Beatty}D.P. Beatty et al.: Phys. Rev. C \textbf{48} (1993) 1428.
\bibitem{Oset} E. Oset and D. Strottman: Phys. Rev. Lett. \textbf{70} (1993) 146.
\bibitem{LoI_S2S} H.~Fujioka et al.: Letter of Intent for J-PARC $50\,\mathrm{GeV}$ Synchrotron
(2016) ``Investigation of Pion Double Charge Exchange Reaction with S-2S Spectrometer''\\
(\url{http://j-parc.jp/researcher/Hadron/en/pac_1607/pdf/LoI_2016-19.pdf})
\bibitem{Kanatsuki} S. Kanatsuki et al.: JPS Conf. Proc. \textbf{8} (2015) 021018.
\bibitem{Nagae} T.~Nagae et al.: Proposal for J-PARC $50\,\mathrm{GeV}$ Proton Synchrotron ``Spectroscopic Study of $\Xi$-Hypernucleus, $^{12}_{\Xi}\mathrm{Be}$, via the $^{12}\mathrm{C}(K^-,K^+)$ Reaction''
\\(\url{http://j-parc.jp/researcher/Hadron/en/pac_0606/pdf/p05-Nagae.pdf}).
\bibitem{LoI_HIHR} H.~Fujioka et al.: Letter of Intent for J-PARC $50\,\mathrm{GeV}$ Synchrotron
(2016) ``Search for tetraneutron by pion double charge exchange reaction on $^4\mathrm{He}$''\\(\url{http://j-parc.jp/researcher/Hadron/en/pac_1607/pdf/LoI_2016-18.pdf})
\bibitem{Sugimura}H.~Sugimura et al.: Phys. Lett. \textbf{B729} (2014) 39. 
\bibitem{Harada}T.~Harada, A.~Umeya, and Y.~Hirabayashi: Phys. Rev. C \textbf{79} (2009) 014603.
\end{thebibliography}
\end{document}